\documentclass[12pt,preprint]{aastex} \def\erf{{\rm erf}} \def\masyr{{\rm
mas}\,{\rm yr}^{-1}} \def\yr{{\rm yr}^{-1}} \def\nltt{{\rm NLTT}}
\def\usno{{\rm USNO}} \def\degsq{{\rm deg}^2}

\begin{document}

\title{Improved Astrometry and Photometry for the Luyten Catalog. I.  Bright
Stars}

\author{Andrew Gould and Samir Salim} 
\affil{The Ohio State University, Department
of Astronomy, 140 W.\ 18th Ave., Columbus, OH 43210} 
\email{samir,gould@astronomy.ohio-state.edu}

\begin{abstract}

We outline the construction of an updated version of the {\it New Luyten
Two-Tenths} (NLTT) catalog of high proper motion stars, which will contain
improved astrometry and photometry for the vast majority of the $\sim 59,000$
stars in NLTT.  The bright end is constructed by matching NLTT stars to
Hipparcos, Tycho-2, and Starnet;  the faint end by matching to USNO-A and
2MASS.  In this first paper, we detail the bright-end matching procedure.  We
show that for the majority of stars in his catalog, Luyten measured positions
accurate to $1''$ even though he recorded his results much more coarsely.
However, there is a long tail of position errors, with one error as large as
$11\degr$.  Proper-motion errors for the stars with small position errors are
$24\,\masyr$ ($1\,\sigma$) but deteriorate to $34\,\masyr$ for stars with
inferior positions.  NLTT is virtually 100\% complete for $V\la 11.5$ and
$|b|>15\degr$, but completeness in this magnitude range falls to $\sim 75\%$ 
at the Galactic plane.
Incompleteness near the plane is not uniform, but is rather concentrated in
the interval $-80\degr<\ell<20\degr$, where the Milky Way is brightest.

\end{abstract}
\keywords{astrometry -- catalogs -- methods: statistical}
%\clearpage
%\newpage
 
\section{Introduction
\label{sec:intro}}

	More than two decades after its final compilation, the {\it New Luyten
Two-Tenths Catalog} (NLTT) of high proper-motion stars %($\mu\geq 180\,\masyr$)
($\mu>0\farcs18\, \yr$)
\citep{luy,1st} and its better known subset, the
{\it Luyten Half-Second Catalog}
\citep[LHS]{lhs} ($\mu>0\farcs5\, \yr$), continue to be a vital source of
astrometric data.  They are still mined for nearby stars
\citep{reidcruz,jahr,scholz,gizreid,henry}, subdwarfs
\citep{gizreid,ryan2,ryan1}, and white dwarfs
\citep{rls,schmidt,ldgs,jones,lwd}. NLTT is at the center of the controversy
over whether halo white dwarfs can contribute significantly to the dark matter
\citep{reidsahu,opp,flynn}; and it is the primary source of candidates for
astrometric microlensing events to be observed by the {\it Space
Interferometry Mission (SIM)} \citep{nearbylens}.  Despite the advent of many
new proper-motion surveys, including Hipparcos \citep{hip}, Tycho-2
\citep{t2}, Starnet \citep{starnet}, UCAC1 \citep{ucac}, as well as deeper but
more localized surveys: SuperCOSMOS Sky Survey in the south \citep{sss1,sss3},
Digital Sky Survey (DSS)-based survey in the Galactic plane \citep{lepine},
search for $\mu > 0\farcs 4 \yr$ stars in $\sim 1400\,\degsq$ \citep{monet},
EROS 2 proper motion search in $\sim 400\, \degsq$ of high latitude fields
\citep{eros}, and MACHO search in $50\, \degsq$ towards the bulge and the LMC
\citep{macho}, NLTT remains unchallenged as a deep, all-sky, proper motion
catalog.

	NLTT is an all-sky position and proper-motion (PPM) catalog with a
proper-motion threshold of $\mu_{\rm lim}=180\,\masyr$.  It extends to $V\sim
19$ over much of the sky, although it is less deep $(V\la 15)$ within about
$10\degr$ of the Galactic plane and also in the celestial south
($\delta<-30\degr$).  In addition to PPM, NLTT lists somewhat crude
photographic photometry in two bands ($B_\nltt$, $R_\nltt$), rough spectral
categories, as well as important notes on some individual stars, primarily
common proper motion (CPM) binaries.

	To be sure, the newer catalogs have superseded NLTT in certain
domains.  By observing a large fraction of the brighter $(V\la 11)$ NLTT
stars, Hipparcos obtained vastly superior astrometry and photometry for about
13\% of NLTT, although in its magnitude limited survey ($V<7.3$--9.0) it did
not find any significant number of new high proper-motion stars not already
catalogued by Luyten.  Tycho-2, which combined re-reduced Tycho observations 
of 2.5 million stars with 144 ground-based catalogs 
(most notably the early 20-century Astrographic Catalog) to derive proper
motions,
includes PPM and photometry for several thousand additional NLTT stars, and
also contains several hundred previously unknown bright $(V\la 11)$ high
proper-motion stars.  However, neither Hipparcos nor Tycho-2 probes anywhere
near the faint $(V\sim 19)$ limit of NLTT.  Moreover, their overlap with NLTT
has never been systematically studied.  UCAC (US Naval Observatory CCD
Astrograph Catalog) is in the process of delivering a new all-sky PPM catalog
down to $R\sim 16$ based on CCD observations.  The first release (UCAC1)
covers 80\% of the Southern hemisphere.  While its photometry is only for a
single band (close to $R$), UCAC1 can easily be matched to 2MASS \citep{2mass}
with its $JHK$ photometry, effectively creating a multi-wavelength PPM
catalog.  Unfortunately, the current release, UCAC1, excludes most of the NLTT
high proper-motion stars, due to absence of these stars in Southern parts of
USNO-A2 catalog \citep{usnoa2}, which was used as a first epoch. (USNO-A2, and
its earlier version USNO-A1, \citealt{usnoa1}, were derived by measuring first
generation Schmidt plates.)\ \ High proper-motion stars pose an especially
difficult problem for automated catalog construction: counterparts of
slow-moving stars can be reliably identified at different epochs, but
fast-moving ones are easily confused with pairs of unmatched, but unrelated
stars, or even with spurious objects.  These problems grow rapidly worse near
the magnitude limit and towards the Galactic plane.  Without either new and
robust (but difficult to write) algorithms or a vast investment of manual
labor, the only routes open in the face of these difficulties are to eliminate
the potential high proper-motion stars from the new catalogs or to include
them but to acknowledge that many may be spurious.  These two approaches have
been respectively adopted by UCAC1 and Starnet. (Starnet derives its proper
motions for 4.3 million stars by combining the \citealt{nesterov} reduction of
Astrographic Catalog with GSC 1.0, \citealt{gsc1}.)

	There are two other PPM catalogs whose release is promised in the near
future, USNO-B \citep{usnob} and GSC-II (\citealt{gsc2}, also known as GSC 2.3
).  Both are based on photographic astrometry and photometry down to or beyond
the NLTT limit of $V\sim 19$. They use a combination of first and second
generation Schmidt plates to determine proper motions.  Neither project has
stated explicitly how they will handle high proper-motion stars.

	Ultimately, space missions such as {\it FAME}, {\it DIVA}, and/or {\it
GAIA} will produce reliable catalogs of high proper-motion stars, since each
star is observed hundreds of times with great astrometric precision, thus
eliminating the possibility of false matching.  However, NLTT stars that are
fainter than the {\it FAME} or {\it DIVA} survey limit of $R\sim 15$ cannot be
observed by them
unless they are securely located prior to the mission \citep{sgo}.
Hence, this is yet another reason for obtaining improved astrometry for these
stars.

	While NLTT has proven incredibly valuable, it also has significant
shortcomings.  As mentioned above, its two-band photometry is relatively
crude, so that classification of stars using the NLTT reduced proper motion
(RPM) diagram (one of the main motivations for constructing the catalog) is
rather uncertain: white dwarfs are easily confused for subdwarfs, as are
subdwarfs for main-sequence stars.  Very few of the NLTT stars have their
optical magnitudes in standard bands available in the literature. This problem
is not easily resolved by matching with other catalogs that have better
photometry, such as USNO-A (photographic $B_\usno$ and $R_\usno$: $\sigma\sim
0.3$) or 2MASS ($JHK$: $\sigma\sim 0.03$). This is because although a large
fraction of NLTT stars have fairly precise ($\sim 10''$) positions, a
significant minority have much larger errors, making automated identification
of counterparts in other catalogs quite difficult.

	Some applications require much higher precision proper motions than
NLTT's characteristic 20--30 $\masyr$.  For example, we \citep{nearbylens}
have previously found that even when faint NLTT stars can be identified in
USNO-A (thus establishing their POSS I 1950s positions to $\sim 250\,$mas) the
NLTT proper-motion errors propagate to create $\sim 1\farcs5$ position errors
in 2010, too large to reliably predict viable astrometric microlensing events
to be observed by {\it SIM}, and so measure precisely the mass of the
lens (high proper motion star). Recently we have found that these errors can
be reduced to $\sim 100\,$mas, which is quite adequate for our purposes, by
matching NLTT stars first to USNO-A and then to 2MASS.  We therefore began to
create our own private catalog of NLTT/USNO-A/2MASS matches.  While this
catalog is highly useful for our specific project, it has many spurious
identifications and is too incomplete to be suitable for general release.

	However, as we have worked with our provisional NLTT/USNO-A/2MASS
catalog, we have gradually become aware of the wide range of potential
applications to which this catalog could be put.  For example, the resulting
$V-J$ RPM diagram (where $V$ is calibrated from $B_\usno$ and $R_\usno$)
permits a much more reliable separation of different classes of stars than did
the original $B_\nltt - R_\nltt$ RPM diagram.  Study of this diagram will be
useful in its own right, but can also guide the interpretation of RPM diagrams
constructed of NLTT stars present in SDSS \citep{sdss}, making use of
excellent SDSS photometry and SDSS/USNO-A derived proper motions.  An updated
NLTT catalog could also be exploited to study binarity as a function of
stellar population and to find new wide-binary companions of NLTT stars.
Finally, it would permit a more thorough study of some of NLTT's own
properties, including most especially its completeness.  NLTT's completeness
(or lack thereof) is central to the debate over the possibility that a
significant fraction of the dark matter can be in halo white dwarfs.
\citet{flynn} argued that if NLTT were assumed to be complete at the bright
end ($V\la 13$), then one could use a bootstrap-type procedure to show that it
is also close to complete at the faint end.  There are two problems with this
argument.  First, it has never been demonstrated that NLTT is in fact complete
at the bright end.  Second, \citet{monet} argued that the bootstrapping
procedure was not valid, at least in its original form.

	Here, we present the first step in the construction of an updated
NLTT catalog: matching bright NLTT entries with those in more recent and 
higher-precision catalogs, primarily Hipparcos and Tycho-2, but also Starnet.
By doing so, we characterize both NLTT's PPM errors and its completeness 
at the bright end for the first time.  These results are of interest in
their own right, but they also serve to guide our search procedure for
USNO-A/2MASS counterparts at fainter magnitudes.

	That faint-end search will be the subject of Paper II of this series.
The general approach will be to match (circa 1950) USNO-A stars with their
(circa 2000) 2MASS counterparts using NLTT (1950 epoch) positions as a rough
guide as to where to find USNO-A counterparts, and NLTT proper motions to
predict the positions of their 2MASS counterparts.  To correctly apply this
approach and to understand its limitations, one must have a good grasp of the
NLTT PPM error distributions, and for this it is best to compare NLTT with
other PPM catalogs.

	The bright-end and faint-end searches are complementary.  On the one
hand, Hip\-par\-cos/Tycho-2/Star\-net become highly incomplete for $V\ga 12$,
and so cannot probe faint magnitudes.  On the other hand, the POSS plates that
were scanned to produce USNO-A saturate for $V\la 11$, leading to increasingly
unreliable photometry, astrometry, and even identifications at bright
magnitudes.  Nevertheless, these searches do have some overlap at intermediate
magnitudes, and we will exploit this overlap to check each method against the
other.  Hence, we push the bright-end search as faint as we can, incorporating
Starnet (with its spurious high-proper motion stars).  And we push the
faint-end search as bright as we can, eliminating only the Hipparcos/NLTT
matches before beginning the search.

	 \citet{bakos} have recently located 96\% of stars from the LHS 
catalog (which contains about 8\% of all NLTT entries).  
They manually identified and measured the positions of LHS stars in DSS1 and 
DSS2 images. The revised coordinates have an
accuracy of $\sim 2''$.  Moreover, now
that most LHS stars (even those with really bad NLTT positions) have been 
located, their positions can be easily further refined by matching them
to other astrometric catalogs, which the
authors did for some bright stars using the Tycho-2 and Hipparcos catalogs.

	In \S\ \ref{sec:asterror}, we analyze the NLTT PPM errors, and in \S\
\ref{sec:strategy} we describe the construction of the revised catalog.  In
\S\ \ref{sec:complete} we discuss NLTT's completeness.  In \S\ 
\ref{sec:NLTTerr}, we study
the RMS differences between proper motions from NLTT and those of the
three modern catalogs.  In \S\ \ref{sec:false}, 
we estimate the rate of false matches of our procedure.  Finally, in
\S\ \ref{sec:bakoscomp}, we perform an additional check on the efficiency
and reliability of our matches by cross-checking with the \citet{bakos}
catalog.  In Paper II we will
merge the results of the bright-end and faint-end searches and make the
resulting catalog publicly available.  This first release of our catalog will
likely take place prior to the full release of 2MASS data, and hence will
include only the 47\% of the sky covered by the 2MASS release.  The catalog
will be updated following the full 2MASS release.

\section{NLTT Astrometric Errors
\label{sec:asterror}}

	To match NLTT stars with counterparts found in other catalogs, we have
to start with NLTT's positions and proper motions.  To devise a search
strategy, it is therefore crucial that we first evaluate the error
distributions of these quantities.  Positions in NLTT are nominally given to
two different levels of precision, indicated by a flag: in right ascension
(RA), either 1 second or 6 seconds of time and in declination (DEC), either
$0\farcm1$ or $1'$ of arc.  The inferior precisions are the rule in the south
($\delta<-45\degr$), and the superior precisions are the rule in the north,
although for $\delta>80\degr$, the RA is given at nominally inferior
precision because 6 seconds of time still corresponds to a relatively small
arc.

\begin{figure}
\plotone{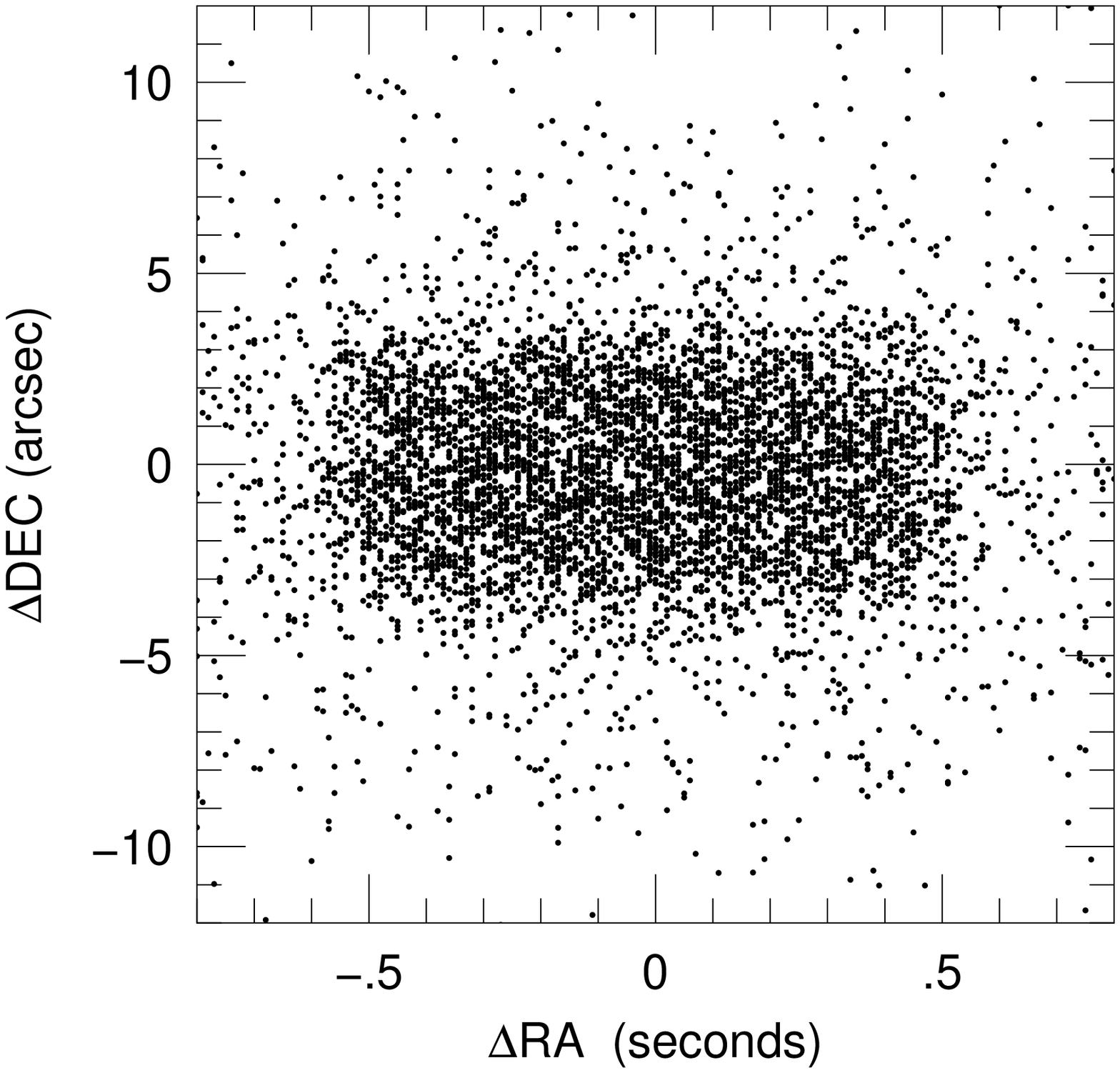}
\caption{\label{fig:pos_rectangle}
Differences between stellar positions as reported in NLTT and the very
accurate positions of the same stars from Tycho-2, both evaluated in the NLTT
epoch of 1950.  The ($1\,\rm s\times 6''$) rectangle that dominates this plot
is caused by the fact that Luyten measured a majority of his stellar positions
to a precision of $1''$, but reported them only to 1 second of time and $6''$
of arc.
}\end{figure}

\begin{figure}
\plotone{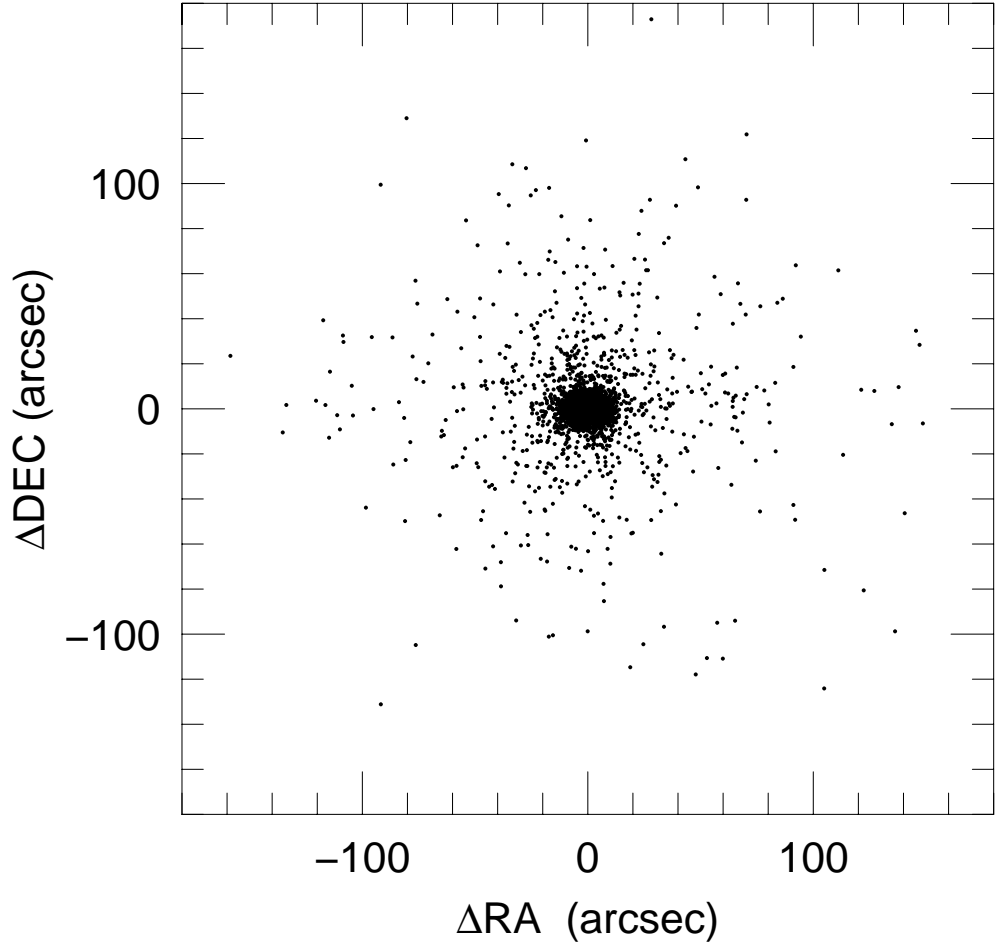}
\caption{\label{fig:pos_circle}
Same as Fig.\ \ref{fig:pos_rectangle} except that first, RA is now plotted in
arcsec rather than in seconds of time, and second, the scale is much larger.
Although most NLTT positions are quite accurate (see Fig.\
\ref{fig:pos_rectangle}), there is a substantial ``halo'' of outliers, some
that go well beyond the dimensions of this plot.
}\end{figure}

	What is the actual PPM error distribution for the $\sim 54,000$ NLTT
entries with nominally good positional precision?  To investigate this, we
first select only Tycho-2 stars with proper motions $\mu>180\,\masyr$, i.e.,
the same limit as of NLTT.  Using Tycho-2 proper motions (having typical
errors of just $2\,\masyr$), we propagate their positions back to 1950, the
epoch of NLTT.  We then find all NLTT entries whose catalog coordinates lie
within $3'$ of each Tycho-2 star -- many times larger than the nominal
positional accuracy.  To ensure a high level of confidence that the counterpart
is real, we restrict consideration to stars for which there is one and only
one Tycho-2/NLTT match, and for which the unique NLTT match does not have a
flag indicating an inferior-precision measurement.  Finally, we eliminate 84
matches for which the Tycho-2 proper motion disagrees with the NLTT proper
motion by more than $75\,\masyr$.  Figures \ref{fig:pos_rectangle} and
\ref{fig:pos_circle} show the position differences (NLTT $-$ Tycho-2) of the
resulting 6660 unique matches, on small and large scales respectively.  The
central portion of this distribution (Fig.\ \ref{fig:pos_rectangle}) exhibits
an obvious rectangle with dimensions ($\rm 1\,s\times 6''$), i.e., exactly the
precisions imposed by the rounding truncation of the catalog entries. Hence,
one can deduce that for the majority of the entries within this rectangle,
Luyten actually measured the positions to much better precision than they were
recorded into NLTT.  Also obvious in this figure is a halo of stars that have
characteristic errors that are of order the discretization noise.  We
therefore model the distribution of Luyten's position measurements as being
composed of two populations, with intrinsic Gaussian measurement errors (in
arcsec) of $\sigma_1$ and $\sigma_2$.  Given the discretization of the
reported positions, the distribution of residuals in DEC $(\Delta \delta)$ 
should therefore be,
\begin{eqnarray}
P(\Delta \delta) & = & {1\over 2w}\biggl\{q\biggl[\erf\biggl({c+ w/2 -
\Delta\delta\over \sigma_1}\biggl) - \erf\biggl({c -w/2 -\Delta\delta\over
\sigma_1}\biggl)\biggr] \nonumber \\ & & +(1-q)\biggl[\erf\biggl({c + w/2 -
\Delta\delta\over \sigma_2}\biggl) - \erf\biggl({c - w/2 -\Delta\delta\over
\sigma_2}\biggl)\biggr] \biggr\},
\label{eqn:ddec}
\end{eqnarray}
where $c$ is the center and $w$ is the width of the
discretization box in DEC and $q$ gives the relative normalizations of the two
populations.  We apply equation (\ref{eqn:ddec}) to the $N=5495$ stars lying in
the strip $7\farcs1 \cos\delta+ 0\farcs5>\Delta \alpha>-7\farcs9
\cos\delta-0\farcs5$, and fit only in the region shown in Figure
\ref{fig:pos_rectangle}, i.e., $|\Delta\delta|<12''$.  (The reason for
offsetting the center of this box from 0 will be made clear in Table 1,
below.)\ \ We find that when $w$ is left as a free parameter, the
best fit value is consistent with $w=6''$, the value expected due to
discretization.  We then fix $w$ and rederive the other parameters.
Next, we write a similar equation for the residuals in RA,
\begin{eqnarray}
P(\Delta\alpha) & = & {\sec\delta\over 2w}\biggl\{q\biggl[\erf\biggl({c + w/2
- \Delta\alpha\over \sigma_1\sec\delta}\biggl) - \erf\biggl({c -w/2
-\Delta\alpha\over \sigma_1\sec\delta}\biggl)\biggr] \nonumber \\ & &
+(1-q)\biggl[\erf\biggl({c + w/2 - \Delta\alpha\over
\sigma_2\sec\delta}\biggl) - \erf\biggl({c - w/2 -\Delta\alpha\over
\sigma_2\sec\delta}\biggl)\biggr] \biggr\}.
\label{eqn:dra}
\end{eqnarray}
We apply equation (\ref{eqn:dra}) to the $N=5022$ stars in the strip
$3\farcs3>\Delta\delta>-3\farcs7$, and fit only in the region
$|\Delta\alpha|<12''\sec\delta$.  The width is again consistent with the
expected value, $w=15''$, so we again hold $w$ fixed.  Table \ref{table:pos}
shows our results.  Here $N$ is the total number of stars in the subsample and
$N_1=Nq$ is the number in the good-precision ($\sigma_1$) population.  For both
the $\alpha$ and $\delta$ directions, we find that $\sigma_1\sim 1\farcs1$,
$\sigma_2\sim 6''$, and $N_1\sim 4000$.  That is, for more than half the full
sample ($4000/6660=60\%$), Luyten actually obtained the stellar positions to a
precision of $1''$ even though he recorded his results much more coarsely.
For most of the rest, his measurement errors were similar to the
discretization noise.  

\placetable{table:pos}

	The offsets $c$ may result from small systematic errors in
Luyten's global astrometry, or from real offsets between his global frame and
the ICRS that underlies Tycho-2 astrometry.  In any case, these offsets are
taken into account when we fix the intervals from which we draw stars to fit
to equations (\ref{eqn:ddec}) and (\ref{eqn:dra}).

	However, Figure \ref{fig:pos_circle} shows that there is an additional
population, an ``outer halo'' beyond what would be predicted by extrapolating
the behavior of the inner two populations.  The structure of the distribution
shown in Figures \ref{fig:pos_rectangle} and \ref{fig:pos_circle} will lead us
in \S\ \ref{sec:strategy} to a ``layered'' approach to identifying NLTT stars.

	Because some stars have much lower position errors than others, we
suspect that they may also have much lower proper-motion errors.  To test
this, we divide the above sample into two subsamples, those lying with a
slightly broadened $(16''\times 8'')$ rectangle and those lying outside it.
For the stars in the rectangle, the RMS difference between the magnitudes of
the proper motion as measured by NLTT and Tycho-2 is $18\,\masyr$, while for
those outside the rectangle it is $24\,\masyr$.  (Tycho-2 error are negligible
in comparison.)\ \ Hence, the rectangle stars indeed have better proper
motions.  For the individual components of the proper motion vector, the
corresponding RMS differences are $22\,\masyr$ and $27\,\masyr$ respectively.
For random uncorrelated errors in RA and DEC, one would expect the magnitude
RMS to equal the component RMS.  The fact that the latter is larger is
probably due to transcription errors in NLTT of the proper-motion direction
(proper motion in NLTT is given as a magnitude and a position angle of
direction).

	While the dispersions of this cleaned sample probably realistically
characterize the intrinsic errors in the NLTT proper motions, they are lower
limits on the RMS differences between NLTT and Tycho-2 for the catalog as a
whole.  This is mainly because binaries (which have been preferentially
excluded from the sample by demanding one-one matches) can cause
proper-motions to differ when measured over different timescales.  We use
these estimates to guide our approach to matching, but will evaluate the
dispersions on the matched sample again in \S\ \ref{sec:NLTTerr}
after carrying out the match.

\section{Strategy to Match Bright NLTT Stars
\label{sec:strategy}}

	We match NLTT sequentially to three proper motion catalogs of bright
stars: Hipparcos, Tycho-2, and Starnet, each in succession containing more
stars, but also generally poorer astrometry. That is, we first
match to Hipparcos.  We then remove from consideration all matched NLTT stars
and match the remainder to Tycho-2 (or rather to the subset of Tycho-2 that is
not associated with Hipparcos stars).  We then repeat the procedure for
Starnet.

	Since NLTT has about 59,000 entries, of which almost 13,000 have
counterparts in these three catalogs, we are especially interested in
developing procedures that can match automatically as many of these as
possible, thereby reducing to a minimum the number that require human
intervention.  For each catalog, we therefore begin by matching stars inside a
$(16''\times 8'')$ rectangle.  The chance that two unrelated high
proper-motion stars will fall so close together is miniscule.  We therefore
place only very weak demands on matches: the magnitude of the vector
proper-motion difference should be less than $100\,\masyr$, and the ``$V$''
magnitudes should agree within 1.5 mag.  For Hipparcos, we use the Johnson $V$
mag given in the catalog.  For Tycho-2, we use the catalog's $V_T$.  For NLTT
we use the ``red'' magnitude $R_\nltt$ (which is actually quite close on
average to Johnson $V$, \citealt{flynn}) except when it is not given, in which
case we adopt $(B_\nltt-1)$ for the ``$V$'' mag.  For Starnet, we use the red
photographic magnitude which ultimately derives from the GSC 1.0.  Of course,
all these various magnitudes are not on exactly the same system, but because
of the inaccuracy of photographic magnitudes, in particular those in NLTT, the
systematic differences are not very important: we are interested only in a
crude discrimination between stars of very different brightnesses.  
We conduct our search only for NLTT stars with $R_\nltt\leq 14.0$.
	
	Singular matches from this rectangle search are accepted without
further review.  There are, however, many multiple matches in both directions.
For example, if one component, say A, of a CPM binary matches to a star in a
given catalog, the other component, B, will almost always match as well.  If
the CPM binary is sufficiently wide to have a separate entry in the catalog
being searched (Hip\-par\-cos/Tycho-2/Star\-net), it may yield 4 ``matches'':
A--A, A--B, B--A, B--B, of which the second and third are false.  All such
multiple matches are investigated, but in many cases they cannot be fully
resolved because the true match to the companion is outside the rectangle or
outside the catalog altogether.  We return to this problem below.  The matches
are then removed from both catalogs, and a second search on the remaining NLTT
stars is then conducted inside a radius $\Delta\theta<120''$ but otherwise
using the same criteria.  These matches are also accepted without review: the
main reason for separating the rectangle and circle searches at this stage is
to reduce the number of multiple-matching candidates.  After resolving double
matches (where possible) we return to the rectangle, but loosen the criteria.
We now demand only that the {\it magnitudes} of the proper motions be
consistent within $80\,\masyr$ but place no constraint on the direction.  This
is to allow for transcription errors in the angle recorded in NLTT.  We also
loosen the tolerance on the agreement in ``$V$'' to 2.5 mag.  The resulting
new matches are then scanned manually.  But again, since the chance that two
unrelated high proper-motions stars will fall in the same rectangle is
extremely small, virtually all of these matches are genuine.  Next, we apply
the same procedure to the $2'$ circle.  All matches are again reviewed by
hand, this time more critically.  If, for example, the angle of the proper
motion and the RA both disagree strongly, but the DEC agrees to within a few
arcsec and the ``$V$'' mags also agree well, we assume that the match is real
and that there are multiple transcription errors.  Of course, we are more
liberal for the regions where NLTT has worse-precision positions.  Finally, we
extend the search to $\Delta\theta<200''$ and with the weak constraints on
proper-motion and magnitude agreement.  We review the results with extreme
caution.  (We do not apply this last step to Starnet because it contains
spurious high-proper motion stars).

	After these automated, and semi-automated searches are completed on
one catalog, we move on to the next.  After they are all complete, we move on
to one final manual search.  In it, we plot the unmatched NLTT, Hipparcos, and
Tycho-2 stars, each using a different color, on a map of the sky using vectors
whose length, orientation, and thickness represent respectively the magnitude
and direction of the proper motion, and the ``$V$'' mag of the star.  This
allows us to find counterparts of NLTT stars with major measurement and/or
transcription errors.  For example, we find three counterparts that disagree
in DEC by exactly $1\degr$ but otherwise are in perfect agreement.  We even
find one that disagrees by exactly $11\degr$ in DEC and about 9 minutes of 
time in RA.  That this entry is a transcription error is obvious from the
name that Luyten assigned to the star ``$-65:2751$'', which corresponds
to its true declination, $\delta=-65^\circ$.

	We then return to the problem of binaries.  We examine every pair of
matches separated by less than $2'$ in NLTT.  In a large fraction of cases,
these are each single matches of well separated stars, but we still check that
we do not have the matches reversed, using the relative separation and
orientation reported in the NLTT notes on CPM binaries.  However, there remain
many multiple NLTT matches to single stars in other catalogs, especially
Hipparcos.  We resolve these whenever possible using the {\it Tycho-2
Double Star Catalog} (TDSC, \citealt{tdsc}) 
which contains PPM and photometry for 
multiple-component objects in Tycho and Hipparcos, both actually associated
multiples and spurious optical doubles.  This catalog also contains many
binaries that are treated by NLTT as single stars.  For some of these,
NLTT contains a note that the entry is actually a binary and gives its 
separation and (usually) the magnitudes of its components.  For others,
NLTT regards the object as a single star.  We make a note of all these cases
for our future work on binaries, but for the present treat all single-entry
NLTT stars as single stars.  Some TDSC entries do not list a proper motion.
For a large fraction of the cases we checked, the positions for these
entries are also significantly in error.  We therefore do not make use of
these entries unless there is corroborating information (from NLTT or 2MASS)
that the positions are correct.  For cases where TDSC does not resolve an
NLTT binary or where the TDSC entry does not contain a proper motion, 
we check to see if the star lies in
the 47\% of the sky covered by the 2MASS release.  If it does, usually 2MASS
resolves the binary and we substitute 2MASS coordinates for those of the other
(e.g., Hipparcos) catalog.  We also note whether we believe that the catalog's
photometry can really be applied to each component (or whether this magnitude
actually refers to a blend or to the other component).  In the latter case, we
adopt ``$V$'' from NLTT rather than the catalog.  For binaries not covered by
TDSC or 2MASS, or for which these catalogs do not resolve the binary, 
we record the two NLTT stars as an unresolved binary.

	Finally, in the spirit of pushing our ``bright'' catalog as faint
as possible for later comparison with the ``faint'' catalog of Paper II,
we make a list of all NLTT CPM binaries for which one
component is matched and whose fainter component has $R_\nltt\leq 14.0$, but is
not matched.  We search for these directly in 2MASS, using the coordinates of
the first component and the separation vector given in the NLTT notes to
predict the position of the second.  We incorporate these by assuming that the
proper motion of the first component is also valid for the second.

\section{NLTT Completeness
\label{sec:complete}}

	We study the completeness of NLTT at the bright end by comparing it to
Hipparcos and Tycho-2.  (Starnet cannot be used for this purpose because it
contains spurious high proper-motion stars).  That is, we count the fraction
of Hip\-par\-cos/Tycho-2 stars that were detected by NLTT as a function of
various parameters.  Before doing this, however, we first ask the opposite
question: what fraction of NLTT stars were detected by Hip\-par\-cos/Tycho-2
as a function of $R_\nltt$ (i.e., NLTT's proxy for ``$V$'')?  The answer to
this question, which is given by Figure \ref{fig:mag_comp}, delineates the
NLTT magnitude range to which our subsequent completeness tests apply.

\subsection{Hipparcos/Tycho-2 Completeness Versus Magnitude
\label{sec:mag_comp}}

	The bold curve in Figure \ref{fig:mag_comp} shows the
Hip\-par\-cos/Tycho-2 completeness as a function of $R_\nltt$, i.e. it is the
ratio of Hip\-par\-cos/Tycho-2 detections (shown by the upper of the two
thin-line histograms) to NLTT detections (bold histogram).  This completeness
falls to 50\% at $R_\nltt=11.6$, as a result of the Tycho-2 magnitude limit.
Hence, our subsequent completeness tests apply approximately to NLTT stars
with $V\la 11.5$.

	Figure \ref{fig:mag_comp} has several other features of note.  First,
there is a peak in NLTT detections at $R_\nltt\sim 8.7$, which is then
reproduced by the histogram of Hip\-par\-cos/Tycho-2 matches, as well as that
of the Hipparcos-only matches just below it.  This turns out to be an artifact
of systematic ``bunching'' of NLTT magnitudes: a histogram of Hipparcos/NLTT
matches as a function of Hipparcos $V$ (not shown in the figure to avoid
clutter) exhibits no such ``premature'' peak, but rather has a single,
relatively broad peak at $V\sim 9.5$.

	Note that Hip\-par\-cos/Tycho-2 completeness is $\sim 100\%$ only to
about $R_\nltt\sim 8.5$, falls to $\sim 95\%$ at $R_\nltt\sim 9.5$, and then
plummets rather sharply.  Given that Hip\-par\-cos/Tycho-2 is itself quite
incomplete significantly below $R_\nltt=11.6$, one might ask whether it can be
used to reliably probe NLTT completeness all the way to this threshold.  It
can be so used if the reasons for NLTT non-detections are independent of the
reasons for Tycho-2 non-detections.  It is clear from the form of the
completeness curve in Figure \ref{fig:mag_comp} that Hip\-par\-cos/Tycho-2
loses sensitivity with faintness.  We will argue below that the NLTT
non-detections are due to crowding, and not due to faintness, since
NLTT goes much fainter than the limits of the modern catalogs.  However,
since crowding can exacerbate problems with detection of fainter objects,
there could be some interplay between these two effects.  We will comment on
the role of this interplay in \S\ \ref{sec:long_comp}.

\subsection{NLTT Completeness Versus Proper Motion
\label{sec:pm_comp}}

\begin{figure}
\plotone{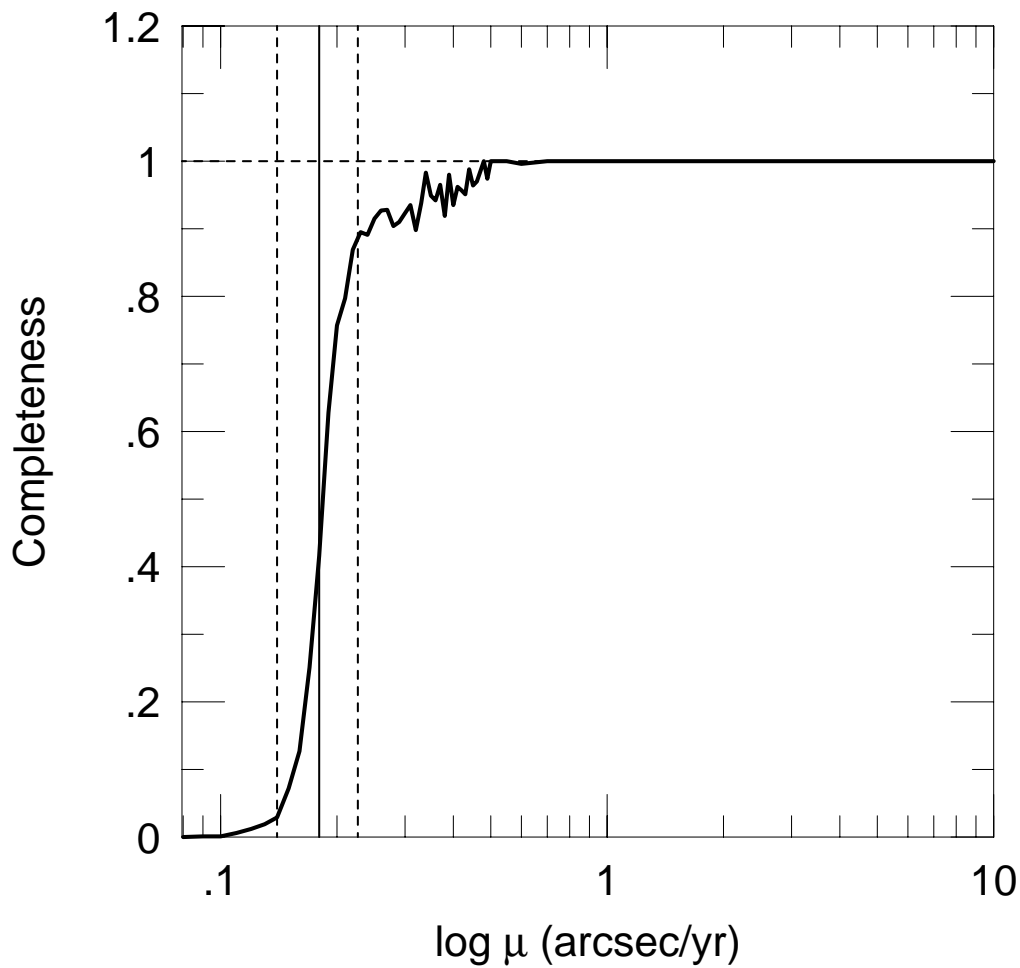}
\caption{\label{fig:pm_comp}
Completeness of NLTT (i.e., the fraction of Hip\-par\-cos/Tycho-2 stars
recovered by NLTT) as a function of Hip\-par\-cos/Tycho-2 proper motion, $\mu$.
The solid vertical line shows the proper-motion limit of NLTT,
$\mu_{\rm lim}=180\,\masyr$, and the two dashed lines show
$\mu_{\rm lim}\pm 40\,\masyr$, i.e., roughly the 1.5 to 2 $\sigma$
errors in NLTT.  The effect of this proper-motion threshold disappears by
$\mu\sim 250\,\masyr$.  Hence, subsequent completeness tests will
be restricted to stars moving faster than this value.
}\end{figure}

	Figure \ref{fig:pm_comp} shows the fraction of Hip\-par\-cos/Tycho-2
recovered by NLTT as a function of proper motion $\mu$ (as measured by
Hipparcos or Tycho-2).  The solid vertical line is at $\mu_{\rm lim}=
180\,\masyr$, the proper motion limit of NLTT.  One expects completeness to
fall by 50\% at this point because half the stars that actually have this
proper motion will scatter to lower values due to measurement error, and so
will be excluded from the NLTT catalog.  One therefore expects completeness to
achieve its asymptotic value a few $\sigma$ above this threshold.  The two
dashed vertical lines are at $\mu_{\rm lim}\pm 40\,\masyr$, which corresponds
to $\pm 2\,\sigma$ for the better-precision NLTT stars, and about $\pm
1.5\,\sigma$ for the others.  The completeness curve does indeed reach an
asymptotic value of $\sim 90\%$ just beyond this point.  Hence, at
$\mu=\mu_{\rm lim}$ we expect the completeness to be $0.5\times 90\% = 45\%$.
The actual value is 42\%.  To avoid this threshold effect, we will restrict
future completeness tests to stars with ``true'' (i.e., Hip\-par\-cos/Tycho-2)
proper motions $\mu>250\,\masyr$.

\subsection{NLTT Completeness Versus Galactic Latitude
\label{sec:sinb_comp}}

	Figure \ref{fig:sinb_comp} (bold curve) shows the fraction of
Hip\-par\-cos/Tycho-2 proper-motion ($\mu>250\,\masyr$) stars recovered by
NLTT as a function of $\sin b$, where $b$ is Galactic latitude.  NLTT is
virtually 100\% complete away from the Galactic plane, but its completeness
falls to about 75\% close to the plane.  This incompleteness is not symmetric:
it is somewhat worse in the south.  We will discuss the reasons for this in
\S\ \ref{sec:long_comp}.  The histogram shows the underlying distribution of
proper-motion stars, from Hip\-par\-cos/Tycho-2, as a function of $\sin b$.
If these stars were distributed uniformly over the sky, then this histogram
would be a horizontal line.  One expects halo stars to be over-represented
near the poles because the reflex of the Sun's motion is most pronounced in
those directions.  One also expects disk stars to be over-represented near the
plane because their density does not fall off with distance in these
directions.  Plausibly, one can see both effects in the Hip\-par\-cos/Tycho-2
histogram.  We will return to this conjecture in Paper II.

\subsection{NLTT Completeness Versus Galactic Longitude
\label{sec:long_comp}}

\begin{figure}
\plotone{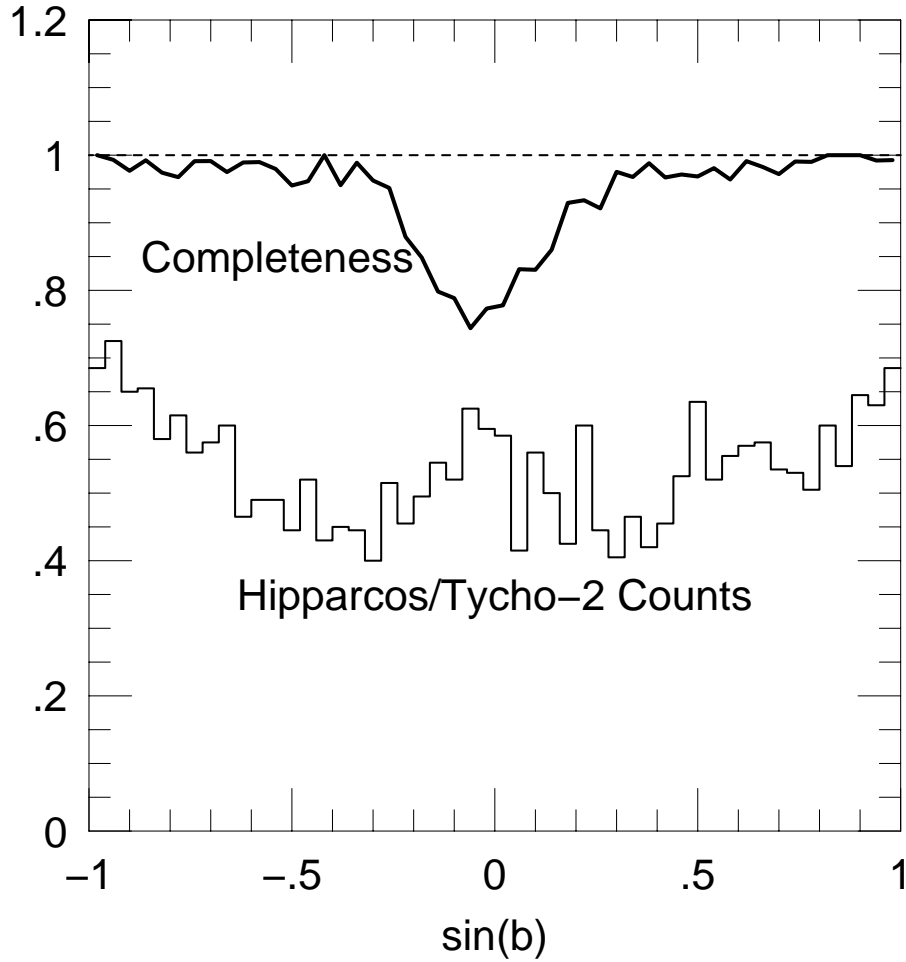}
\caption{\label{fig:sinb_comp}  
Completeness of NLTT ({\it bold curve}) as a function of $\sin b$ where $b$ is
Galactic latitude, i.e., the fraction of stars with $\mu>250\,\masyr$ in
Hipparcos and Tycho-2 (whose distribution is shown by the histogram) that are
recovered in NLTT.  (The histogram has been divided by 200.  The bin size is
0.04)\ \ Incompleteness is significant only close to the plane, where it is
somewhat skewed toward the south.
}\end{figure}

\begin{figure}
\plotone{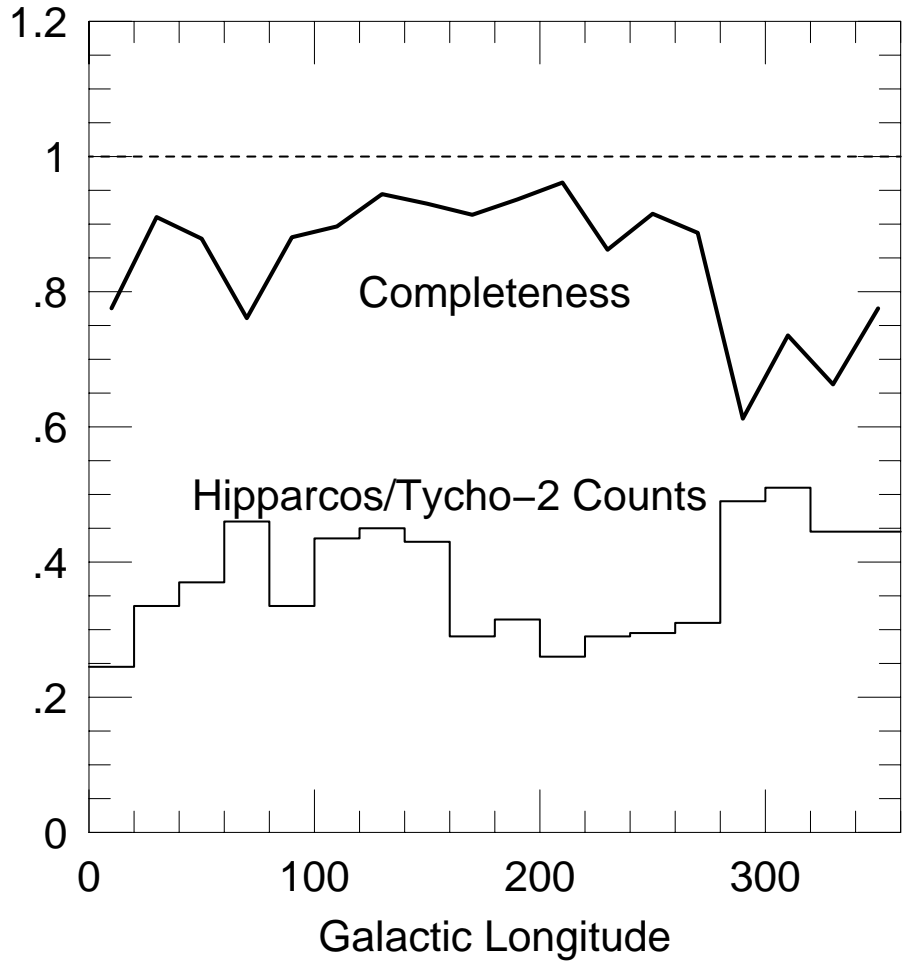}
\caption{\label{fig:long_comp}
Completeness of NLTT ({\it bold curve}) as a function of Galactic longitude
for the subset of stars lying close to the plane $(|b|<15\degr)$.  Also shown
is a histogram (counts divided by 200, $20\degr$ bins) of
Hipparcos and Tycho-2 stars with $\mu>250\,\masyr$ and $|b|<15\degr$.  The
correlation (or lack thereof) between the two curves is analyzed in Fig.\
\ref{fig:long_corr}.
}\end{figure}

	Figure \ref{fig:long_comp} (bold curve) shows the fraction of
Hip\-par\-cos/Tycho-2 proper-motion stars lying in the Galactic plane
($|b|<15\degr$) that are recovered by NLTT as a function of Galactic
longitude.  While the curve is somewhat noisy, there is a clear increase in
incompleteness over the interval $-80\degr\la\ell\la 20\degr$.  This is the
brightest contiguous region of the Milky Way, which lends credence to the idea
that NLTT incompleteness is traceable to crowding-induced confusion. The areas
just south of the Galactic equator are on average brighter than the
corresponding areas just to the north, so the asymmetric behavior seen in
Figure \ref{fig:sinb_comp} also lends credence to this hypothesis.

	In \S\ \ref{sec:mag_comp}, we entertained the possibility that
detection failures in NLTT and Hip\-par\-cos/Tycho-2 might be correlated
which, if it were the case, would undermine the completeness estimates
obtained from the fraction of Hip\-par\-cos/Tycho-2 stars recovered in NLTT.
Figure \ref{fig:long_comp} shows that this effect cannot be very strong, if it
exists at all.
	
\begin{figure}
\plotone{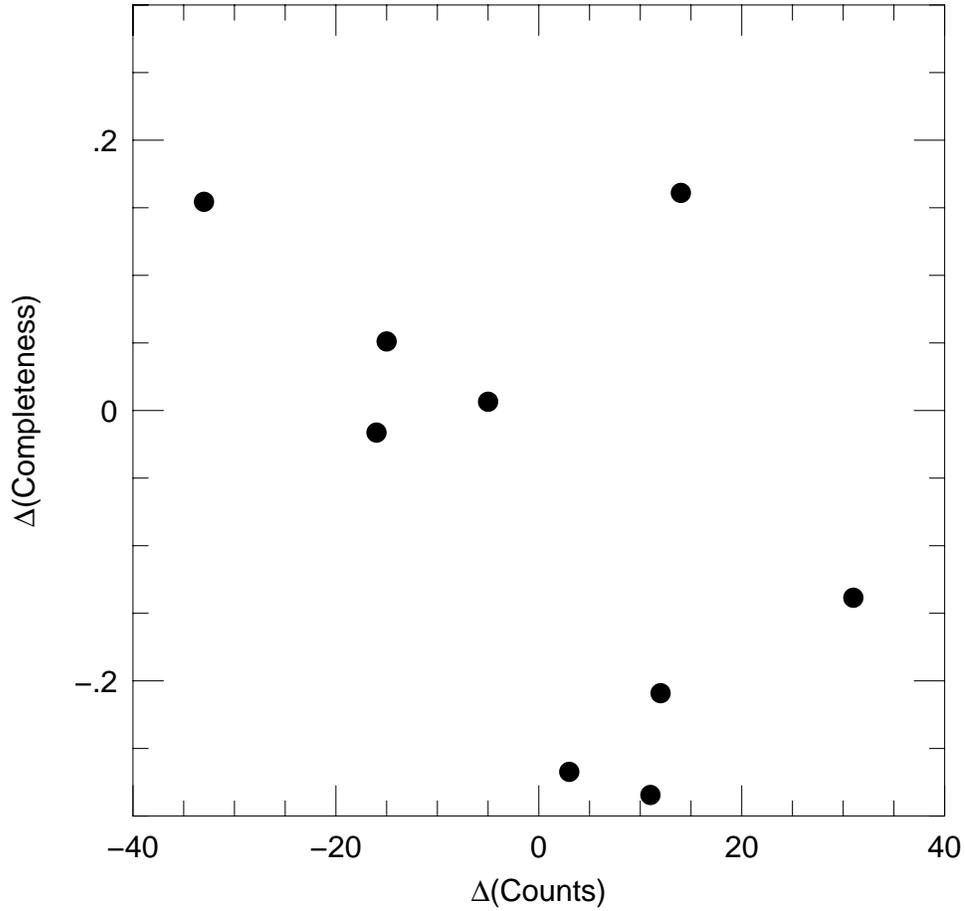}
\caption{\label{fig:long_corr}
Differences in completeness at antipodal points $(\ell\leftrightarrow
\ell+180\degr)$ versus differences in counts of Hip\-par\-cos/Tycho-2 stars
restricted to $\mu>250\,\masyr$ and $|b|<15\degr$ for the same longitude 
pairs.  If Hip\-par\-cos/Tycho-2 completeness were correlated with NLTT
completeness, one would expect these points to be positively correlated.  If
anything, they are weakly anti-correlated. 
}\end{figure}

	The expected number of high proper-motion stars as a function of
Galactic longitude need not be uniform, and will in general depend on the
model of the Galaxy.  However, in any plausible model, the number should be
the same looking in directions separated by $180\degr$ because NLTT stars are
not at sufficiently large distances to probe the Galactic density gradients.
Consequently, the distribution is a result of bulk kinematic effects, which
should be identical in antipodal directions.  Hence, if there
were a correlation, one would expect that pairs of antipodal points with a
positive difference in NLTT completeness would also have a positive difference
in Hip\-par\-cos/Tycho-2 counts.  No such pattern is seen in Figure
\ref{fig:long_comp}.  To make this point clearer, we plot these quantities
against each other in Figure \ref{fig:long_corr}.  If anything, there appears
to be a weak anti-correlation.

	In any event, the primary implication of Figure \ref{fig:sinb_comp},
namely that NLTT is essentially 100\% complete away from the Galactic plane
remains true, independent of these more subtle considerations.  In Paper II,
we will develop a more sophisticated version of the \citet{flynn} bootstrapping
technique to extend this completeness analysis to fainter magnitudes.

\section{NLTT Proper Motion Errors
\label{sec:NLTTerr}}

	Altogether, we have matched 12,736 stars from NLTT to Hipparcos,
Tycho-2, and Starnet (or in a few cases, to CPM companions of these stars that
we found in 2MASS).  We began this study by estimating the errors in positions
and proper motions of the original NLTT catalog, but restricted to a
relatively clean subset.  Here we give the RMS differences between NLTT proper
motions and those found in the three more modern catalogs.  These differences
can result from NLTT measurement errors or transcription errors, from real
differences in the proper motion due to binarity, from misidentification of
counterparts, or from proper-motion errors in the three modern catalogs. 

The last of these four causes can be quite significant for the 
relative handful of stars near 
the magnitude limits of these catalogs, particularly Hipparcos.  Indeed,
we find that when Hipparcos reports errors larger than $10\,\masyr$, its
true errors can be much larger than the tabulated errors.  We also find
that for the sample of stars common to NLTT, Hipparcos, and Tycho-2, the RMS
tabulated Tycho-2 errors are substantially smaller than those of Hipparcos.
Moreover, since these are established by a longer baseline of observation,
they are more directly comparable to the NLTT proper motions in cases
where the Hipparcos proper motion may be corrupted by a short-period 
($P\la 5\,\rm yr$) binary.  We therefore, first substitute Tycho-2 for
Hipparcos proper motions whenever the former are available.  For purposes
of comparing with NLTT we then eliminate all stars with tabulated proper-motion
errors greater than $10\,\masyr$ in either direction.  This removes 52
(0.6\%) of Hipparcos stars, no Tycho-2 stars, and 23 (1.7\%) of Starnet stars.
After removal of these 75 stars, we expect that the proper-motion differences
between NLTT and the three modern catalogs are dominated by the first three 
causes listed above.

	  Table \ref{table:pm} lists these RMS values for various
subsets of the catalog.  Here ``Hipparcos'' refers to NLTT stars found in
Hipparcos, ``Tycho-2'' to stars found in Tycho-2 but not Hipparcos, and
``Starnet'' to stars found in Starnet, but not in either of the other two
catalogs.  ``Better precision'' stars are those with positions specified in
NLTT to 1 second of time and $6''$ of arc (and generally lying in the range
$-45\degr<\delta<80\degr$), and ``worse precision'' stars are the remainder.
When two NLTT stars, usually close components of a CPM binary, are matched to a
single entry in the PPM catalogs, we compare the proper-motion measurements
only once.

\placetable{table:pm}

	Table \ref{table:pm} shows that among the stars with nominally better
positions, those inside the rectangle have consistently lower proper motion
errors than those outside the rectangle.  For Hipparcos and Tycho-2, the stars
with nominally worse positions are intermediate in their proper motion errors
between these two categories, indicating that they are also probably a mixture
of intrinsically better and worse precision measurements.  However, for
Starnet, the proper motion errors of the stars outside the rectangle as well
as those with nominally worse position errors both have significantly worse
proper motion errors.  This may indicate that these stars suffer many more
false matches.  We conduct a preliminary test of this hypothesis in the next
section, although a full investigation will have to wait for Paper II.

\section{False Matching Rate
\label{sec:false}}

	There is an important potential channel for false matches in the work
reported here: If none of the 3 PPM catalogs contain a
counterpart to a given NLTT star, then our search procedure, which looks
farther and farther afield, and with ever weakening demands on agreement in
magnitude and proper motion, may ultimately find a marginally consistent
candidate match, which is nonetheless false.  In fact, since we search the
catalogs sequentially, we could even find a false match to, e.g., a Tycho-2
star, when the true counterpart of the NLTT star was in Starnet.  Ultimately,
in our full catalog, we expect to be able to correct the great majority of
such false matches by checking the results of our bright-end search against
those of the faint-end search that will be performed in Paper II.

	For purposes of the present paper, in which we are presenting only
statistical results and not the catalog itself, it is sufficient to measure
the false-matching rate rather than finding every single false match.  We can
then check whether the rate of false matches is high enough to corrupt the
statistics.

	To this end, we conduct a preliminary version of the faint-end search
that will be fully described in Paper II.  We find all USNO-A stars lying
within a $16''\times 8''$ rectangle of each NLTT star, and then propagate
these forward to the epoch of 2MASS using the NLTT proper motion.  We then
find all 2MASS stars lying within $5''$ of these predicted positions.  We
apply this procedure to the approximately 1/3 of the sky covered by the
overlap of POSS I and 2MASS.  (As already mentioned, high proper-motion stars
are systematically removed from USNO-A in the regions south of POSS I.)  We
find from this exercise that about 3/4 of all NLTT stars in this region lie
within the rectangle.  That is, we obtain an independent check on $3/4\times
1/3 = 1/4$ of all stars.  Because of its very small size, the probability of a
false match within the rectangle is extremely low.

We exclude from this search only the NLTT stars that were matched to Hipparcos
in the first two search stages.  Since these had rather strict criteria, and
since the Hipparcos high proper-motion stars were largely drawn directly from
NLTT, it is very unlikely that there are false matches among these.  We
manually inspect all discrepant matches.  We find no false matches among the
remaining Hipparcos stars, four false matches among the Tycho-2 stars, and
eight false matches among the Starnet stars.  In addition, we find that four
of the Starnet matches, while probably correct, have proper motions that are
seriously in error.  We thus extrapolate that roughly 65 of the more than
12,000 bright NLTT stars that we have matched to Hipparcos, Tycho-2, and
Starnet, are in fact incorrectly matched.  Since this is well under 1\%, it is
unlikely to affect our overall statistical results in a major way.  However,
since roughly $4\times (8+4)/(435+195)\sim 8\%$ of all Starnet matches lying
outside the rectangle (or having inferior positions) are false matches or have
seriously incorrect proper motions, false matching could be responsible for
the exceptionally large proper-motion discrepancies reported for these stars
in Table \ref{table:pm}.  Further investigation will be required in Paper II.

\section{{Comparison With Bakos et al.}
\label{sec:bakoscomp}}

	We can gain some additional sense of the reliability and
efficiency
of our procedure by comparing our results to those of \citet{bakos},
who manually identified the great majority of LHS stars and measured their
PPMs from digitized plates (DSS1 and DSS2).  
As elsewhere in this paper, we restrict
consideration to stars with $R_\nltt\leq 14.0$, since the modern PPM
catalogs to which we are matching NLTT contain essentially no stars
beyond this limit.

\begin{figure}
\plotone{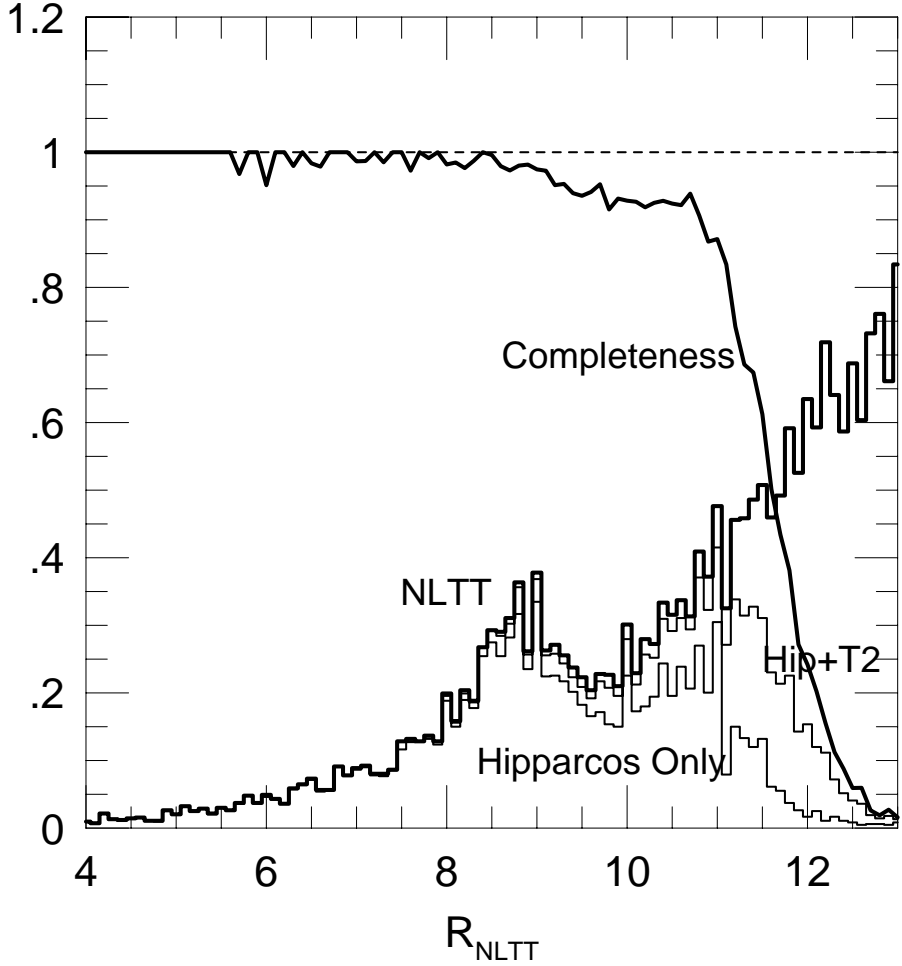}
\caption{\label{fig:mag_comp}
The completeness of the combined Hipparcos and Tycho-2 catalogs as
a function of $R_\nltt$ (roughly Johnson $V$) magnitude, measured
from the fraction of NLTT stars ({\it bold histogram}) that are matched
to one of these two catalogs (upper {\it thin-line histogram}).  Also shown
are the Hipparcos-only matches.  The ``bump'' in NLTT detections at
$R_\nltt\sim 8.7$ is an artifact of NLTT mags.  See text.  Completeness
falls to 50\% at $R_\nltt = 11.6$.  Hence, the subsequent tests on
completeness of NLTT apply directly to its brighter stars, $V\la 11.5$.
(All histograms are divided by 1000.  The bin size is 0.1 mag.)
}\end{figure}

	There are a total of 2682 LHS stars contained in NLTT that meet this
criterion, of which we match 1253.  The vast majority of the rest 
have $R_\nltt \geq 11.0$, at which magnitudes
Hipparcos/Tycho-2/Starnet are rapidly becoming incomplete (see Fig.\
\ref{fig:mag_comp}), and so many nonmatches are expected.  There
are a total of 6 bright nonmatches with $R_\nltt<10.0$ and an additional
14 with $10.0\leq R_\nltt < 11.0$.  All 20 were detected by \citet{bakos} 
directly on the plates.  We therefore use the \citet{bakos} coordinates to 
search
for these stars in the PPM catalogs.  Of the 6 in the brighter magnitude
interval, 5 are in TDSC, but since either the primary or the secondary
does not have proper-motion information and therefore their physical binarity
could not be established, they were excluded from our
identification procedure.  The sixth (LHS 537) is not in any PPM
catalog.  Of the
14 in the fainter interval, one is in TDSC but was excluded for lack of
proper-motion information, and the remaining 13 are not in any catalog.
We conclude that our procedure is missing very few stars that are
accessible through our catalogs, at least among the faster (LHS) stars.

	Besides efficiency, we also use \citet{bakos} to test the reliability 
of our matches.  Of our 1253 matches, \citet{bakos} 
failed to find 32.  Of these 32, we judge 31
of our matches to be highly reliable based on the close agreement
between the NLTT PPMs and the Hipparcos, Tycho-2 or Starnet
PPMs that we recover.
One match, a Starnet star, appears dubious.  Half of
the \citet{bakos} nonmatches are due to
transcription errors in the LHS positions, typically of order $1\degr$,
which evidently prevented \citet{bakos} from finding the correct star.
All but one of these LHS transcription errors were corrected in NLTT.
(On the other hand, there were also new transcription errors present in NLTT
but not LHS, including the star ``$-65:2751$'' [LHS 3674] mentioned in
\S\ \ref{sec:strategy}, which has an $11\degr$ error in NLTT but is
approximately correct in LHS.)\ \  All of the remaining 16 \citet{bakos}
nonmatches for which we obtain good matches (13 from  Hipparcos, 1
from Tycho-2, 2 from Starnet) lie within $135''$ of their predicted NLTT
positions.  However, for these cases DSS2 data were either unavailable to
\citet{bakos} or were of poor quality.

	We search for discrepancies among the 1221 matches that are common
to \citet{bakos} and us.  In the great majority of discrepant cases, there is
good agreement between the Hipparcos/Tycho/Starnet PPMs and those of NLTT,
leading us to believe that our identifications and measurements are
correct.  In most of these cases, \citet{bakos} have a note indicating
that saturation or blending caused difficulties in their measurement.
Less frequently, LHS position errors led \citet{bakos} to measure the
wrong star (recognizable from its negligible proper motion). A bug in
the \citet{bakos} computer program caused them to output a positive DEC
when the star lay within $1\degr$ south of the equator, which accounted
for about a half dozen of the discrepancies.
In one case we traced the discrepancy to our error: we reversed
the identifications of an equal-brightness binary relative to the
convention
of NLTT.

	These tests confirm that our false matching rate is well under 1\%
and that we are missing much less than 1\% of the stars that can be identified
with the catalogs we are using.

\acknowledgments We thank S.\ Frink for providing access to the Starnet
catalog, which has been made available to the SIMBAD database
by S.\ R{\" o}ser but is not yet accessible there. We acknowledge the use
of CD-ROM versions of USNO-A1.0 and A2.0 catalogs provided to us by D.\
Monet.
This publication makes use of catalogs from the Astronomical Data
Center at NASA Goddard Space Flight Center, VizieR and SIMBAD databases
operated at CDS, Strasbourg, France, and data products from the Two
Micron All Sky Survey, which is a
joint project of the University of Massachusetts and the IPAC/Caltech,
funded by the NASA and
the NSF.This work was supported by JPL contract 1226901.

\clearpage

%\begin{figure}
%\plotone{mag.ps}
%\caption{\label{fig:mag_comp}
%}\end{figure}

\clearpage
\begin{deluxetable}{l r r r r r r}
\tabletypesize{\footnotesize}
\tablecaption{NLTT Position Precisions \label{table:pos}}
\tablewidth{0pt}
\tablehead{
% fit      sigma_1    sigma_2    c       q   N    N_1
\colhead{fit} &
\colhead{$\sigma_1$}   &
\colhead{$\sigma_2$}   &
\colhead{$c$} &
\colhead{$q$}  &
\colhead{$N$} &
\colhead{$N_1$} \\
}
\startdata
RA & $1\farcs 1$ & $7\farcs 0$ & 0.03s 
& 0.774 & 5022 & 3887 \\
DEC & $1\farcs 1$ & $5\farcs 8$ & $0\farcs 15$
& 0.735 & 5495 & 4040 \\
\enddata
\end{deluxetable}

\begin{deluxetable}{l l l r r r r }
\tabletypesize{\footnotesize}
\tablecaption{NLTT Proper Motion Precisions \label{table:pm}}
\tablewidth{0pt}
\tablehead{
\colhead{Catalog} &
\colhead{Rectangle}   &
\colhead{Position}   &
\colhead{Number}  &
\colhead{$\sigma (|\mu|$ } &
\colhead{$\sigma (\mu_{\alpha})$ } &
\colhead{$\sigma (\mu_{\delta})$ } \\
\colhead{} &
\colhead{} &
\colhead{quality} &
\colhead{} &
\colhead{$\masyr$} &
\colhead{$\masyr$} &
\colhead{$\masyr$} \\
}
\startdata
 Hipparcos &  In   &    Better  &    5261  &  20  &  22 &  25 \\
           &  Out  &    Better  &    1623  &  31  &  33 &  32 \\
           &  All  &    Worse   &    1284  &  24  &  28 &  34 \\
 Tycho-2   &  In   &    Better  &    1470  &  25  &  24 &  24 \\
           &  Out  &    Better  &    1005  &  32  &  35 &  35 \\
           &  All  &    Worse   &     438  &  28  &  32 &  42 \\
 Starnet   &  In   &    Better  &     738  &  25  &  27 &  24 \\
           &  Out  &    Better  &     435  &  44  &  55 &  65 \\
           &  All  &    Worse   &     195  &  42  &  57 &  45 \\
\enddata
\end{deluxetable}

\end{document}